# Many-to-many. Usability challenges of entity reconciliation in art history and photographic studies

## Abstract

**Purpose**
This article investigates challenges in reconciling heterogeneous records across cultural institutions, focusing on art historical photo archives within the PHAROS consortium.

**Design/methodology/approach**
Through case studies, the study analyses reconciliation workflows and cataloguing traditions, with attention to institutional contexts, data granularities, and modelling strategies.

**Findings**
Reconciliation is seldom a one-to-one operation. Ambiguities, incomplete data, shifting attributions, and varying practices shape outcomes. Strategies observed include the creation of anonymous or collective entities, the use of umbrella terms, the addition of uncertainty qualifiers, and reticence when ambiguity cannot be resolved.

**Originality**
The article highlights the need to model uncertainty explicitly, offering a framework that connects technical reconciliation methods with institutional practices.

**Practical implications**
Insights from PHAROS provide guidance for designing more robust, interoperable, and sustainable cultural heritage infrastructures.

## 1. Introduction[1]

In the Galleries, Libraries, Archives, and Museums (GLAM) sector, Linked Data has emerged as a central strategy for building large-scale metadata aggregators that support cross-collection discovery and research (Hyvönen, 2012; Isaac et al. 2012). By publishing data as Linked Open Data and aligning it to shared authority resources such as VIAF, Getty vocabularies, or Wikidata, institutions can expose their collections in a way that makes them interoperable and reusable beyond local catalogues. The creation of such aggregators depends critically on data reconciliation, to match URIs describing the same people, places, or artworks that are often duplicated, fragmented, or expressed differently across institutions. Reconciling these entities is therefore essential to ensure that aggregators present coherent, enriched, and usable networks of cultural heritage information.

---

[1] 

However, to the best of our knowledge, despite the proliferation of reconciliation methods (Zeng et al. 2021) and empirical studies in the cultural heritage domain, there remains a notable lack of documented processes and policies for reconciling data between cultural institutions. Previous work has largely focused on technical approaches but the organizational and policy dimensions of cross-institution reconciliation are underexplored. While it is difficult to envision a single universal framework given the specificities of institutional cataloguing practices, many of the issues surfaced in comparative analyses of institutional data are recurrent and shared across contexts. Documenting these analyses is therefore crucial, as they can inform the design of reconciliation policies and best practices that will support future large-scale initiatives in the arts and humanities.

This article examines the lack of clear policies for reconciling entities within artwork records. Drawing on the collaborative work of the PHAROS association of art historical photo archives, we report on efforts to reconcile records for artists, photographers, and artworks, highlighting the challenges encountered and the solutions developed by the association. Findings include effective strategies that can be generalised to Cultural Heritage and Humanities domains, which face similar issues, often on the same pieces of information. The framework proposed—integrating identification, modelling, workflow, and interface strategies—therefore offers a reusable structure for any GLAM organisation seeking to consolidate heterogeneous datasets under a shared, semantically interoperable environment.

## 2. State of the art

Galleries, libraries, archives, and museums typically start reconciliation tasks considering the authority control that best fits their data scope. Authority control ensures standardised forms of names, titles, and subjects can be used as coherent access points to collections in organisation and discovery tools (Joudrey et al. 2015). Controlled forms on names are collected in authority files. VIAF clusters national authority files from libraries that describe authors, artists and historical figures. The Getty Vocabularies (ULAN/AAT/TGN) expose Linked Open Data about artists, materials, artwork types, and historical places, that many institutions reconcile to. Nowadays, Wikidata is also an optimal candidate for entity reconciliation and authority matching, since the wealth of annotations on prosopographical and geographical data include direct links to major cultural institutional identifiers and third-party authorities such as the aforementioned VIAF and Getty vocabularies.

Ontologies standardize what is being reconciled. CIDOC-CRM (CIDOC 2021) provides the core event-centric model for describing Cultural Heritage data. Museum aggregators adopt application profiles such as Linked Art (Delmas-Glass and Sanderson, 2020), which offers an Application Programming Interface (API)[2] that simplifies cross-institution alignment and downstream reconciliation. Similarly, Europeana's Entity API exposes canonical records of people, places, and subjects for linking[3]. CIDOC-CRM is increasingly used in combination with the IFLA Library Reference Model, whose object-oriented formulation (LRMoo) has been integrated into CIDOC-CRM to harmonise bibliographic and museum documentation practices (Riva and Žumer 2018).

In practice, semi-automatic curation of alignments is a standard procedure, e.g. using OpenRefine or Getty services (Delpeuch 2019), often combined with local rules for data normalization, variant handling, management of birth/death dates, roles or occupations, etc. When datasets are published as RDF (Baker et al. 2011), frameworks such as LIMES (Ngomo et al. 2021) let curators declare link specifications (properties, similarity metrics, thresholds) and run efficient matchers with pruning rules for large graphs. Supervised Machine Learning and, increasingly, deep learning augment (not replace) matching rules (Di Cicco et al. 2019). These techniques are applicable to humanities corpora when paired with domain constraints and authorities.

Nonetheless, in the GLAM context—where names, attributions, and historical identities are noisy—curator oversight is essential. OpenRefine Reconciliation API underpins many authority-mapping workflows but has recognized scoring/UX limitations in need of improvement (Knoblock et al. 2017; Delpeuch 2019). Even with strong authorities and methods, there are many-to-many realities, including homonyms, shifting attributions (sure artist vs. attributed), uncertain life dates, role and legal status granularity (photographer vs. studio), which experts themselves would struggle to clearly identify (Doerr 2025). Notably, experts from cultural heritage institutions involved in reconciliation are willing to invest significant time and effort to ensure that the final data is accurate (Knoblock et al. 2017). However, very few institutions published strategies for managing ambiguity (clustering, evidence aggregation, periodic reprocessing), which exemplify the need for policies and processes, not just algorithms (Hickey and Toves 2014). For instance, the treatment of variant identities in RDA (Resource Description and Access), especially the "work group" construct, which aggregates heterogeneous manifestations under a common conceptual umbrella when provenance or authorship cannot be uniquely determined, is a notable attempt to solve compelling identification and reconciliation issues (Coyle and Hillmann 2007). Likewise, the modelling of collections and subcollections by *Analytical Model of Collections and their Catalogues (*Heaney 2000*)*, later formalised in the Dublin Core Collection Description Application Profile (DC 2024), sheds light on the importance of data modelling as a proxy for exposing ambiguities in formal languages. Overall, given the peculiarities of the documentation held by PHAROS institutions and photographic collections themselves, the reconciliation challenges encountered in PHAROS are not specific to photographic archives but reflect structural issues across libraries, archives, and museums, where descriptive traditions and authority sources evolve independently and require principled alignment strategies.

---

[2] https://linked.art/about/1.0/
[3] https://europeana.atlassian.net/wiki/spaces/EF/pages/2324561923

## 3. Background

The PHAROS association includes 13 art historical photo archives from Europe and North America. Their collections together amount to 31 million photographic documents depicting artworks from all periods (Caraffa et al. 2020). Seven institutions (Bibliotheca Hertziana – Max Planck Institute for Art History, Bildarchiv Foto Marburg, Frick Art Research Library, Fondazione Federico Zeri, Kunsthistorisches Institut in Florenz, Netherlands Institute for Art History (RKD), Warburg Institute) are currently piloting the online research platform artresearch.net, which to date includes 1.7 million photographs of about 1 million artworks.

The platform artresearch.net is a collaborative initiative built on ResearchSpace (Oldman and Tanase 2018), enabling the ingestion of data transformed into Linked Open Data. To ensure interoperability, the PHAROS association has adopted an application profile of CIDOC-CRM as its standard to harmonise metadata descriptions and offer, where possible, a unified view of the collections. The reconciliation activities are an integral part of the Extraction/Transformation/Loading (ETL) process of artresearch.net. Such activities are still in progress, due to the continuous ingestion of new data sources and the consequent revision of policies for data management (e.g. extension of the ontology, controlled vocabularies, and matching rules).

Original data leverage several metadata standards to describe photographic documentation of artworks, such as the Italian ICCD national cataloguing standards for describing artworks (Scheda OA)[4] and photographs (Scheda F)[5], the Marburger Informations-, Dokumentations- und Administrations-System (MIDAS) metadata standard[6], MARC21 (Library of Congress 2021), and institutional standards, all addressing a core set of similar metadata fields for the description of artworks (e.g. artists, material and techniques, dates, subjects, keepers, provenance) and photographic documentation (e.g. photographers, techniques, shot and printing dates, copyright holders).

The data described in art historical photo archives is deeply heterogeneous, shaped not only by the diversity of the photographic objects themselves but also by the cataloguing practices that have framed their curation. Cataloguing has historically been an expensive and labor-intensive process, subject to shifting institutional priorities, disciplinary interests, and technological limitations. In many archives, early cataloguing efforts reflect the biases of the moment: choices about what to describe, how to describe it, and which granularity to prioritize were conditioned by the knowledge frameworks of the time. In particular, in cataloguing campaigns launched in the 1990s, the description focused solely on the reproduced objects, while little attention was paid to information related to the photographic medium, often recorded merely as captions or free-text descriptions. As a result, the metadata in photo archives is often uneven, partial, or inconsistent.

---

[4] http://www.iccd.beniculturali.it/it/ricercanormative/29/oa-opere-oggetti-d-arte-3_00
[5] http://www.iccd.beniculturali.it/it/ricercanormative/62/f-fotografia-4_00
[6] https://d-nb.info/1212592956/34

From the 2000s onward, as systematic cataloguing of photographic collections gained momentum and sensitivity towards photographic sources raised, archivists worked without established authority files or standardized vocabularies. While attempts were made to create shared terminologies across institutions, these efforts were still emerging and lacked the robustness later achieved in fields such as art history.

Indeed, in the art history domain, scholarly interest and methodological advances led to improved cataloguing practices. The revision of cataloguing practices in art historical photo archives were driven by the shifting interests in the scholarly and art market communities, which increasingly requested more rigorous and coherent access to data collections, e.g. by artist or genre, which required specialised intervention to improve data quality. Despite these efforts, data often retained idiosyncrasies born from local traditions and cataloguers' individual judgments.

The rise of Semantic Web technologies offered an unprecedented opportunity to revisit past cataloguing practices and to address long-standing terminological and interoperability challenges. However, the very same technologies expose the fragility of the underlying data, marked by uncertainty, vagueness, and inconsistencies, inherited from earlier cataloguing and by the impossibility to uniquely and incontestably identify entities (e.g. artists' followers). As a result, records cannot always be aligned in a one-to-one manner with external authority files or even reconciled across institutions. A comprehensive revision of legacy data is prohibitively costly, and, likely, it wouldn't solve uncertainties inherited from divergent scholarships (e.g. unknown artists differently identified by scholars).

For photo archives data, the challenges are magnified. As mentioned earlier, photographic studies as a field emerged far later than art historical studies, and only relatively recently has the photograph been recognized as a research object in its own right. The absence of comprehensive terminologies for photographs and photographers has left significant gaps in the descriptive frameworks available to institutions. Unlike artworks, for which authority control and controlled vocabularies gradually developed, photographs continue to suffer from a lack of standardized descriptors. This absence not only complicates cross-institutional interoperability but also risks perpetuating the unevenness and biases embedded in earlier cataloguing efforts. Consequently, metadata, now of great importance to scholars, must be historicized and reinterpreted in the light of a previously unsolicited demand for precision.

Cataloguers, metadata specialists, and developers are left to grapple with this heterogeneity and the inherent impossibility of achieving precise reconciliation. Moreover, the level of precision demanded by structured data systems stands in tension with the epistemic uncertainties of art history itself. Any transformation process must therefore acknowledge, and design around, the impossibility of achieving absolute accuracy.

## 4. Challenges in content data reconciliation

PHAROS data sources selected for ingestion and reconciliation reveal the strong imprint of institutional biases and cataloguing traditions. These manifest in varying levels of accuracy when describing the same pieces of information—an outcome of shifting priorities and interests in cataloguing over time—as well as in differing levels of detail, reflecting changing expectations of data quality and content. Taken together, these patterns echo several known phenomena in cataloguing processes (Barabucci et al. 2022; Vitali and Pasqual 2025), namely:

- Reticence. Information is not recorded due to haste or lack of policies.
- Flattening. A metadata field is used to describe different pieces of information at the same time.
- Coercion. Information for which no appropriate field is found, is forced into other fields.
- Dumping. Information for which no appropriate field is found, is placed as plain text inside a descriptive field.
- Varying reliability. Not all pieces of information available for an artefact have the same reliability, but motivations, sources and acknowledgement of uncertainty may be missing.
- Assumed certainty. Even when there is no disagreement on the value of a field, there could be different levels of confidence about the truth of that value.

Such issues may lead to inconsistent access points to records (flattening, reticence), prevent automatic processing of data and entities (coercion, dumping), and the inability to distinguish between subjective and factual information

(varying reliability, assumed certainty), for which the reader is implicitly led to assume factuality and certainty of the records.

Designing the artresearch.net platform, and therefore the reconciliation process, requires a constant compromise between representing uncertainty and ensuring usability. While it is crucial to preserve and communicate the varying levels of certainty and reliability embedded in the data, exposing such granularity directly to end users risks overwhelming them and undermining the clarity of the interface. Scholars and general users alike may find themselves disoriented if uncertainty is foregrounded as a first-class feature of the user experience (Battisti and Daquino 2025). For this reason, usability concerns inevitably guide design decisions. Information must be presented in a way that remains accessible and meaningful, without masking the underlying complexity but also without burdening the user with excessive interpretive labor. In PHAROS we adopted multiple solutions over time, sometimes testing several approaches in parallel in order to evaluate results (as presented in Section 5).

However, difficulties in reconciliation stem not only from biases introduced by the cataloguing process, but also from the intrinsic features of the data itself. For example, the same photographer may appear under different name variants across institutions—sometimes with incomplete dates, sometimes with conflicting attributions—which makes automated alignment with authority files or across data sources particularly challenging. We present here a selection of such reconciliation issues that required the intervention of metadata specialists and the co-design of policies with institutions to establish consistent practices, negotiate acceptable compromises in the User Interface (UI), and ensure that divergent cataloguing traditions could be aligned without erasing their historical specificities.

## *Artists*

Artists in PHAROS are primarily reconciled against ULAN, the most authoritative and complete list of artists of all times, and secondarily against Wikidata when no ULAN match exists. In practice, however, straightforward reconciliation is often hampered by complex cases that require policies to be defined in advance to guide the review of candidate matches. We highlight three recurring situations.

**Attribution qualifiers.** Many records attribute works to anonymous artists whose identities are framed in relation to a known figure—for example "Leonardo da Vinci, school", "follower of", or "circle of". While maintaining the connection to the known artist is important for retrieval, in PHAROS such cases are modeled as distinct anonymous entities related to the identified artist. This approach allows us to distinguish between known and unknown individuals—previously "flattened" into the same field—while still preserving the associative relationship. These relationships can then be exposed to users of artresearch.net through hierarchical facets.

**Family names.** Archivists often recorded only a family name when a specific attribution was not possible, or when the artwork was the result of a family collaboration. The label "Bellini", for instance, might refer to Jacopo, Giovanni, Gentile (all of them being artists from the same family), to a subset of them, to the family collectively, or to an otherwise unidentified relative. Interpretations also vary across institutions: in one collection "Bellini" may consistently mean Giovanni, while in another it may be used inconsistently across different records. To address this ambiguity, we create a distinct entity for the anonymous group represented by the string "Bellini", without asserting formal relationships to identified family members. The link remains only at the level of use interface, in facets, where anonymous groups and identified individuals can be visually juxtaposed, thus respecting the reticence already present in the original cataloguing.

**Diverging identities.** In some cases, pseudonymous attributions have been identified differently by scholars. For example, "Pseudo Ambrogio di Baldese (1375–1425)" has been variously associated with Ventura di Moro (1399–1486, ULAN 500012920) and Lippo d'Andrea (1371–1451, ULAN 500082343). Since ULAN itself does not endorse a single identification—listing the pseudonym as an alternative name for both candidates—PHAROS creates a new URI for Pseudo Ambrogio di Baldese on which institutions can agree. This URI is linked to both ULAN candidates using rdfs:seeAlso, rather than the semantically stronger owl:sameAs, thereby encoding alignment while preserving uncertainty. Artists' names are presented in user interfaces along with all links to ULAN URIs. Institutional endorsements of one attribution over another are neither recorded nor shown in the interface, since disagreements can exist even within a single institution. For instance, when archives originate from one specific art historian, such as Federico Zeri, archivists are expected to respect the creator's hypotheses without reassessment.

A parallel case is Ercole Grandi, whom ULAN identifies with Ercole de' Roberti (ULAN 500124891), while some institutions treat them as distinct individuals. To avoid endorsing either position while preventing duplicates in the PHAROS authority file, we apply the same strategy: a new URI is minted for Ercole Grandi, linked to the ULAN candidate via rdfs:seeAlso. Again, institutional endorsements are not preserved, ensuring neutrality where scholarly consensus is lacking.

**Anonymous groups**. When an artist or a group cannot be identified, neither with a name or pseudonym, cataloguers record it as an anonymous entity. Some institutions may operationalise conventions to name anonymous entities, for instance recording the school, the region, and the century, e.g. "Anonymous Florentine 16th century". The same label is used across records despite not being possible to distinguish whether and when the label refers to the same anonymous person or group. Therefore, the reconciliation of such records inside the same institution, let alone across institutions, presents inconsistencies at semantic level. In PHAROS, for the sake of usability purposes, this granularity is overlooked and one entity only is created to represent anonymous groups with the same space-time frame.

## Materials, techniques, artwork types

One of the most striking reconciliation oddities arises from terms that simultaneously imply different conceptual levels that appear separated in the AAT, such as materials, techniques, and work types. For instance, the term "albumen" used in a photo record may refer to the raw material itself (AAT 300011802), the photographic process employing albumen (AAT 300133274), or to the specific product "albumen print" (AAT 300127121), which in turn is often conflated with the broader category of "paper". Such terminological overlaps complicate reconciliation, since a single string can ambiguously point to three distinct entities in a controlled vocabulary. This creates inconsistencies both across and within institutions: cataloguers may choose one AAT term over another when reconciling their terms. As a result, automated alignment to authority records risks introducing semantic noise unless supported by policies that explicitly disambiguate these layered concepts. In PHAROS we opt for the most comprehensive approach, asking cataloguers to identify terms for all conceptual levels potentially involved in the definition of the term.

## Provenance: keepers and locations

The reconciliation of current and previous artwork keepers presents complications at modelling and reconciliation levels. In principle, a keeper is an agent—either a person or an organisation—that may be associated with a residence, itself housed in a building located in a specific city.

**Agents or buildings.** Ambiguities first emerge in the interpretation of the semantic nature of an entity: is a museum to be considered an organisation (a conceptual agent) or a building (a physical object)? Automatic reconciliation to Wikidata often reveals inconsistencies, since Wikidata contributors have not always followed a uniform rationale—sometimes registering a cultural institution as an organisation, other times as its building, or both (e.g. Q3330350). In PHAROS we standardise this by aligning to organisational entities, as they better account for the possibility of relocation over time.

**Collections as agents and hybrid structures.** Another example of flattening in the metadata field is when keepers are recorded in ways that do not respect a coherent semantic distinction. For example, collections are often recorded as keepers, even though they are groupings of objects rather than agents capable of acquiring or holding them. To handle these cases, PHAROS reconciles entities representing the owners of collections whenever this can be univocally identified, alternatively minting a URI for an anonymous owner. Moreover, museums and the collections they preserve are often recorded interchangeably, e.g. "Frick Art Research Library" and "Frick Collection". Again, we reconcile the institution that is accounted as keeper rather than the collection, defining modelling solutions to bind the artwork to the specific collection.

Alongside collections, PHAROS data include a heterogeneous range of entities identified as keepers: auction houses (agents), art dealers (agents), museums (agents), and buildings (physical objects). These discrepancies produce peculiar hierarchical arrangements that require careful reconciliation to avoid misrepresenting localisation. Cases where collections, buildings, and institutions are combined in layered descriptions are particularly challenging, as the flattening of semantic roles obscures the distinction between the agents who act, the places where they reside, and the objects they contain. Examples include movable and immovable entities, such as buildings part of another building ("Inferior basilica, Basilica of S. Clemente"), museums in a building ("Museum of Porziuncola, Basilica of S. Maria degli Angeli"), museums in another museum ("The Fine Arts Museums of San Francisco, M.H. de Young Memorial Museum"), collections in a museum ("Ajmanov grad, collection J. Demsar"), collections in a building ("Allington Castle, collection Lord W.M. Conway"), auction houses in a building ("Hotel George V, Auction Tajan"), and antiquarians in a building ("The Manor House, Phillips of Hitchin (Antiques) Ltd"). The reconciliation of each individual entity happens respecting the semantic nature of the entities involved: an action house will be reconciled to its equivalent agent in Wikidata, and the place where it resides to the respective administrative location. To preserve the relation between the two entities as necessary condition to properly locate the artwork, we adopt modelling solutions that bind movable entities (e.g. collections) to immovable ones (e.g. a building) with respect to the artwork at hand.

Manors, such as Broomford Manor or Chatsworth House, are a further example of semantic flattening, this time stemming from their inherently ambiguous administrative nature. They sit somewhere between place names and buildings, making it difficult to unambiguously identify the administrative region or town to which they belong. This ambiguity is reinforced by the fact that the names of collections preserved in manors often coincide with the names of the manors themselves, blurring the semantics of the keeping entity. We adopt the same approach of Wikidata, whenever applicable, treating them primarily as places.

**Last known locations**. Recording an auction house or an antiquarian as a keeper usually implies that the artwork was last seen in their hands at a certain date (noted either in the same or in another metadata field). However, the artwork most likely moved elsewhere afterwards. While reconciliation addresses these agents (auction houses and antiquarians) as stable entities, the event being recorded is actually a temporary observation of the artefact at a particular place and time—something conceptually different from being permanently kept by an institution.

**Granularity of places.** Reconciling geographical entities—such as the current location of a museum within a city or town—introduces another level of complexity. Wikidata offers comprehensive data for geodata reconciliation, but automatic alignments often produce uneven hierarchies of geographic specificity. For example, the Wikidata geolocation of the National Gallery of Art in London identifies its headquarters as the City of Westminster (which is part of London), a level of granularity too specific to display to end users and requiring post-processing. In PHAROS, we accept these automatic alignments but apply harmonisation rules to normalise places to the city level.

**Provenance reconstruction.** All these issues make reconstructing the sequence of provenance events especially challenging. Merging descriptions from different institutions that use inconsistent levels of granularity (e.g. referencing a person instead of a collection, or vice versa), and that often contain missing steps or contradictions, requires extensive work on data normalisation, conflict resolution, and the management of vagueness and uncertainty. Moreover, the community lacks a shared authority of art provenance events that could be used to reconcile sales, acquisitions, and changes of ownership. To further complicate matters, such information is frequently recorded only as free-text descriptions (a case of dumping), using editorial conventions that vary widely across institutions, making automatic extraction of entities and event sequences difficult.

Currently, PHAROS adopts an incremental approach to delivering this information to end users. While the knowledge extraction and reconciliation processes are still ongoing, the textual provenance descriptions are displayed as-is in artresearch.net records, allowing users to access all the available information from different providers in one place—even if it has not yet been fully interpreted, deduplicated, or merged.

## Artworks

Despite the significant progress reported by GLAM institutions in the production of Linked Open Data (LOD), reconciling artworks to a shared and comprehensive authority list remains a challenging task. Getty CONA can be considered a potential venue that unfortunately has not received broad attention like other Getty standards, and no such comprehensive list currently exists in LOD. Consequently, PHAROS focuses on reconciling artworks across its partner institutions, enabling a coherent and unified view of their catalogues within artresearch.net. This reconciliation, which can be recognised as one of the main objectives of the process, relies on a combination of strategies, primarily involving metadata alignment and visual similarity analysis.

**Metadata (mis)alignment.** Metadata-based reconciliation alone presents substantial challenges, especially for artworks described using different conventions in entries containing competing, contradictory, or inaccurate information. For example, artworks may have been referenced under different titles over time, and photo archives may either adopt variant historical titles or supply their own when none were available, introducing potential mismatches. Dates can also be recorded at different levels of granularity (centuries, decades, year ranges, or vague estimates), and attributions can vary from one institution to another. The same issues apply to artists: one institution may list an anonymous member of a Master's workshop, while another may name the actual identified artist who was part of that workshop. These discrepancies are widespread and make metadata alignment alone an unreliable method for comprehensive reconciliation.

**Visual similarity.** To address this, PHAROS first identifies candidate matches through image-similarity algorithms, and only then compares and reviews the associated metadata with the involvement of art historians and cataloguers. This approach, however, introduces its own difficulties. For instance, exact copies produced by copyists are often matched by the algorithm and might appear correct when only metadata are considered, potentially misleading non-experts into interpreting the result as evidence of differing scholarly opinions. Only expert visual comparison can confirm or reject such matches. In these cases, while a strict equivalence link would be incorrect, an associative relation should still be maintained. Similarly, preparatory drawings or individual components of composite artworks may be matched to their completed versions. Here too, a one-to-one reconciliation would be misleading, while preserving associative relationships remains valuable.

## Photographers

Photographers in PHAROS can be people, ateliers or cultural institutions commissioning the photograph. These are reconciled to Wikidata whenever applicable. The decision to reconcile against a general knowledge base rather than a specialised one derives from the variety of photographs provenance represented in PHAROS, for which specialised authority lists would only cover small groups (e.g. Photographers' Identities Catalogue (PIC)[7] mostly cover American photographers), requiring us to adopt several ones. Wikidata instead offers a broader scope, allowing us a higher coverage, even if it lacks completeness for more niche professionals.

Data about photographers—primarily names and dates—are usually derived from the transcription of inscriptions or stamps found on photographs. As noted earlier, early cataloguing campaigns often recorded these names without an explicit intention to identify a stable entity that could later serve as a data access point. In such cases, photographers' names functioned more as captions than as authority data. Over time, some institutions, such as the Zeri Photo Archive, revised and simplified these naming conventions to produce their own authority files, but most others did not. Likewise, a few institutions adopting the MIDAS standard have created an internal unified authority list, though it is not comprehensively interlinked to external sources. This has resulted in significant variation across collections, with divergent naming conventions, attribution practices, and levels of accuracy. Compared with the reconciliation of artists to ULAN, the reconciliation of photographers in PHAROS poses challenges that stem largely from the inherited vagueness and instability of names and the solutions devised are driven by the need of simplifying facets for accessing data. These issues can be summarized as follows.

**Uncertain legal status associated with a name.** A recurring issue arises when a name may refer either to a person or to a studio, without sufficient evidence to disambiguate. For example, *"Böhm"* may designate the photographer *Osvaldo Böhm* or his studio *Foto Böhm*. While institutions often record one of the two forms—clarifying the legal status—this is not always possible. In ambiguous cases, the string *"Böhm"* (sometimes followed by a "?") is used without distinction, making it impossible, either programmatically or empirically, to determine the intended entity. To handle these cases, we create a new entity labelled "Böhm (collective name)" that represents an anonymous group. Similarly to the above scenario of artists' family names, we do not materialise relations between the uncertain entity "Böhm (collective name)" and the certain entities "Osvaldo Böhm" and "Foto Böhm", because of their blurred boundaries. Instead, we mint yet a new URI representing the commonly used label "Böhm" and we link it to all the aforementioned entities. We informally call this new entity "umbrella term", which does not represent a real-world entity (such as a person or a group), rather, it is a reified label that we use to collect and mildly relate entities under a common string. In the user interface, *"Böhm"* appears as a top-level facet, expandable into the underlying certain ("Osvaldo Böhm" and "Foto Böhm") and uncertain ("Böhm (collective name)") entities.

---

[7] https://pic.nypl.org/

**Changes in constituent aspects of an entity.** Photographic studios or ateliers, like other organizations, often change their names, locations, or ownership, raising questions about whether these should be modeled as distinct entities or merely as name variants. Wikidata reflects this inconsistency by offering URIs at different levels of granularity: sometimes one URI covers multiple manifestations of a company, while in other cases each manifestation has its own record. For example, Braun & Cie (WD Q79493769) subsumes entities such as Maison Adolphe Braun & Cie and Braun, Clément et Cie, treating them as alternative forms of the same company. In PHAROS, however, each manifestation found in the data is associated with its own newly minted URI, with reconciliation established only to the Wikidata entity that best matches. An additional umbrella term is then created to group all these manifestations under a single label. Another example concerns the company "Armoni Moretti Raffaelli". The studio was opened by Luigi Armoni and then inherited by Luigi Raffaelli, changing name in "Armoni e Raffaelli", and later by Mario Moretti, changing name in "Armoni Raffaelli & Moretti". In Wikidata there is an imprecise record describing the photographer "Luigi Armoni Raffaelli", identified as a person. In PHAROS, we mint separate URIs for each institutional form and then aggregate them under an umbrella term labeled "Armoni Raffaelli Moretti".

**Changes of family name**. Individual name changes also complicate reconciliation. Hilde Lotz-Bauer, a photographer and art historian, is recorded in PHAROS under multiple names reflecting marital status and institutional conventions: "Lotz-Bauer, Hilde", "Degenhart-Bauer, Hilde", and "Bauer Lotz, H.". While reconciliation to her Wikidata entry (WD Q1618235) poses no semantic problems, the user interface can display only one canonical form, thus obscuring the temporal and biographical significance of these variations. To preserve this information, we employ CIDOC-CRM to create relationships between photographs and the linguistic objects that capture the exact forms recorded by cataloguers, which are then related back to the person entity.

**Institutions commissioning photographs**. When the photographer is unknown but the commissioning institution—often acting as the copyright holder—is recorded, cataloguers frequently entered the institution's name in the same metadata field (another case of coercion). In PHAROS we chose a simplified modelling approach: rather than multiplying anonymous entities, we retain the institution within the authority list of photographers for usability purposes. For example, the Rijksmuseum in Amsterdam (WD Q190804) may appear both as a photographer and as the custodian of an artwork. Special cases arise when a subdivision or related office is named, such as the Fotocommissie Rijksmuseum Amsterdam. Since such entities lack dedicated Wikidata records and cannot be reconciled with the broader institution, PHAROS creates separate entities grouped under the umbrella term "Rijksmuseum". No direct relation is asserted between these entities, as their legal status may have shifted over time. This preserves the necessary vagueness while simplifying the user interface facets.

However, this simplification faces clear limitations when both the institution and the photographer are known. Should the individual alone be recognised as a photographer, or should the institution also be acknowledged? While reconciling such entities independently is straightforward, managing them within a single authority list of photographers can introduce inaccuracies. For example, Mario Sansoni worked both independently and for several ateliers and institutions over time (e.g., Alinari, the Frick Art Research, Nesti, Bencini). Recording only Sansoni as a photographer, while discarding information about the institutions that employed him, fails to capture the evolution of his professional career and the specific context in which photographs were produced. Addressing this issue requires extensive modelling work to represent time-sensitive collaborations between individuals and institutions, thereby enriching the authority list with semantic relations.

**Multiple or alternative photographers.** In PHAROS data, we sometimes find entries such as "Anderson/Brogi." Here, two photographers—or their respective ateliers—are recorded together, but it is unclear whether this indicates an actual collaboration or alternative attributions. Once again, the issue of uncertain legal status arises, requiring the use of anonymous entities and "umbrella terms" to represent the ambiguity. At the same time, the relation between the photo and its possible authors (one, both, or either) raises modelling challenges in how to represent uncertainty in the interpretation of the string and, consequently, in the attribution of authorship, which is not endorsed by cataloguers. In PHAROS we adopt custom ontology patterns to represent alternative statements and named graphs to preserve their provenance.

**Uncertain names.** Beyond uncertainty about the legal status of an entity represented by a string, we also encountered cases where ambiguity affects the very semantics of reconciliation. For example, the label "Beyer, Constantin?" suggests that the photographer may indeed be Constantin Beyer (WD Q95218985), but could also refer to a different, unrelated individual. In such cases, we reconcile the string to the identified Wikidata record and explicitly model the uncertainty of the attribution at the level of the specific photograph. Other cases are more problematic—for instance, "Beyer, ?", where it is impossible to determine whether the reference is to Constantin, Günther, or Klaus Beyer. Here, since no reasonable guess can be made, we create an anonymous entity unlinked to any of the candidates, deliberately avoiding attributions and thus remaining reticent in suggesting one or more possible identifications.

# 5. Challenges in Managing Ambiguity and Heterogeneity in Cultural Heritage Data

The case study reveals that uncertainty in cultural heritage data is not a marginal exception but a structural condition of cataloguing practices. Based on the challenges observed in PHAROS, we propose a four-part framework for managing uncertainty that integrates modelling, workflow, and interface considerations. This framework makes explicit the steps required to reconcile heterogeneous records belonging to different institutions, while preserving provenance and institutional perspectives.

1. **Identification Strategies.** Methods for detecting when entities differ subtly or substantially, including homonyms, pseudonyms, anonymous groups, script variations, and incomplete historical attributions. These strategies rely on both authority data (e.g., VIAF, ULAN, Wikidata) and institutional metadata to flag potential ambiguities before automated reconciliation takes place.
2. **Modelling Strategies.** A set of representational techniques used to encode the ambiguity rather than suppress it. These include:
    a. distinguishing between possible and certain equivalences (using rdfs:seeAlso rather than owl:sameAs when needed);
    b. describing associative relations when the above similarity/equivalence relations cannot be defined between related entities;
    c. employing higher-level conceptual nodes (e.g., umbrella terms) to group heterogeneous or conflicting attributions without erasing institutional differences;
3. **Workflow Strategies.** Multi-step processes combining automated reconciliation, authority-based matching, conflict detection, and human validation. The framework emphasises division of labour across institutions, enabling each archive to review candidates for its own materials while sharing merged results in a centralised graph.
4. **Interface and Interaction Strategies.** Decisions about how uncertainty is conveyed to end users, including visual grouping of variant titles, suppressing overly granular distinctions when they hinder navigation, and presenting provenance trails to show how each statement entered the graph. These strategies treat user experience as a constitutive part of reconciliation, not an afterthought.

This framework is generalisable beyond photographic archives, offering conceptual tools applicable to museums, libraries, and archives facing similar challenges in cross-institutional metadata alignment.

## 5.1 Naming and Identity Ambiguity (People, Groups, Pseudonyms)

The overall strategy of PHAROS is to align entity names (people, organisations, places) to third-party authorities whenever applicable, minting new URIs only when a match cannot be found, and later looking for a cross-institutional reconciliation. Authority datasets such as ULAN, VIAF, GND, and Wikidata improve coverage but also introduce competing conventions, further complicating identity resolution.

Our approach combines automated similarity measures with authority cross-checking to generate candidate matches, which are then validated by institutional experts. When uncertainty persists, relationships are encoded either 1) using non-committal predicates (e.g., rdfs:seeAlso) instead of owl:sameAs (e.g. for artists), or 2) using umbrella terms to define associative relations between real-world entities and a shared label.

The first case applies to artists, which are reconciled to ULAN and Wikidata through a two-step process combining rule-based algorithms and graph embeddings, with final verification by human reviewers. Our initial attempt used large language models (LLMs, specifically Gemini 2.5 Pro) to perform automatic reconciliation. We provided the LLM with structured data on artists from various institutions (including names and life dates) alongside a list of partially name-matched ULAN candidates, asking it to discard those with low string similarity or incompatible dates. However, this approach produced unpredictable errors and hallucinations—such as incomplete outputs, missing fields (e.g., dates), duplicates, false negatives (missing existing matches), and false positives (proposing matches not in the input list)—which were difficult to diagnose or correct.

We therefore shifted to a rule-based approach that compares string similarity and dates. Although this method has low recall, it produces highly precise matches, enabling a manageable first round of human review without overwhelming reviewers with likely incorrect matches. For all remaining unresolved cases, we apply a second method based on graph embeddings. This approach leverages names and dates from local catalogues as well as biographical data from ULAN. Here, LLMs are used in a more constrained role: they prune the candidate matches suggested by the embedding model and return the most likely matches for human review. This combined method yields more reliable results in cases where the rule-based system cannot confidently identify a match. However, discrepancies arise when the same artist described in different institutions is differently matched to ULAN candidates. For instance, one institution may use the full name of the artist "Gavasio, Giovanni Giacomo", which gets a correct match in ULAN, while another institution uses a different form of the name, "Gavazzi Giovanni", that does not find a confident match in ULAN. In these cases, the expert review of candidate matches allows us to prevent duplications in the final list, eventually retrieving the correct ULAN match if available.

Notably, some institutions already provide reconciliation to existing authorities. The Frick Art Research Library includes links to the Library of Congress (LOC) artists' IDs, the Bibliotheca Hertziana includes links to GND and Wikidata, and the Zeri photo archive includes links to ULAN, VIAF and Wikidata for a subset of entities. To harmonise reconciliations and retrieve ULAN matches for all artists, we perform a cross-authority reconciliation, automatically retrieving links between LOC, GND, VIAF, ULAN and Wikidata records. Unexpectedly we find many situations where datasets somehow disagree with each other. For instance, a Wikidata URI is linked to a URI in VIAF. In turn, that very same VIAF URI points to a different wikidata URI. In PHAROS, such cases amount to 27% of cross-reconciled URIs. In Fig. 1 we exemplify our workflow to harmonise cross-authority links.

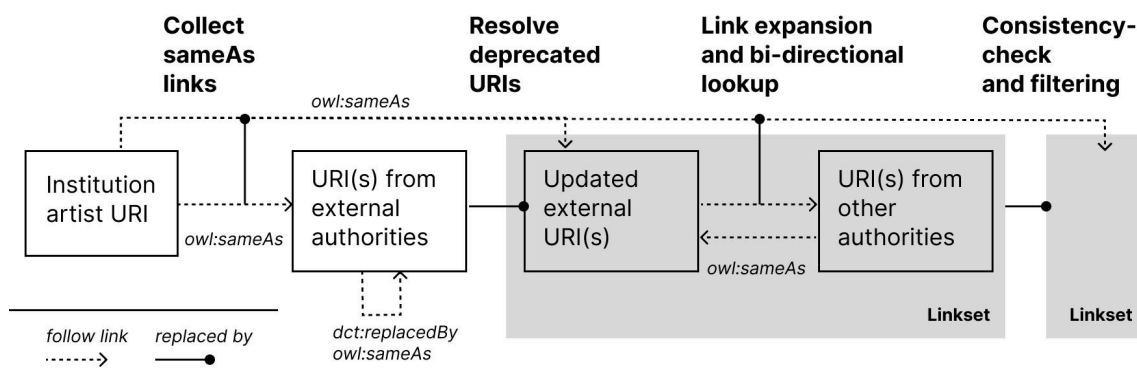

Figure 1. Cross-authority reconciliation process

In detail, we begin by collecting all sameAs (or equivalent) links provided by each institution for the entity in question, and we first resolve any deprecated or obsolete authority URIs. For example, in the Zeri photo archive we follow links to ULAN and Wikidata for artists, checking dct:replacedBy in AAT/ULAN and owl:sameAs in Wikidata, and replacing outdated URIs where necessary. We then expand each linkset as much as possible. For instance, the Frick Art Reference Library provides links to Library of Congress (LoC) records. We follow the LoC link to determine whether the corresponding LoC record contains further links to Wikidata, ULAN, and other relevant authorities. We also perform the reverse check: if the Wikidata record contains a link back to the same LoC URI. Through this process, each entity from an institution is associated with a comprehensive linkset of sameAs connections, each annotated with its provenance. Next, we assess whether these linksets are internally consistent. Ideally, exactMatch/sameAs links should behave transitively—for example, LoC URI → Wikidata URI → LoC URI should return to the original identifier. In practice, this is frequently not the case: an LoC URI (A) may point to a Wikidata URI (B), while B in turn points to a different LoC URI (C). When such inconsistencies occur, we filter out the conflicting reconciliations. To select the most reliable mappings, we prioritise the Library of Congress, Deutsche Nationalbibliothek, RKD Artists, and ULAN as the most authoritative sources, using Wikidata as a fallback given its active editorial community. We deliberately exclude aggregators such as VIAF from reconciliation decisions, as their records are updated less frequently and therefore are more likely to contain outdated or broken external links.

Another example is the conflict of matches inherited from cross-reconciliation performed at institutional level. For instance, the Zeri photo archive provides reconciliations for "Bazzi Giovanni Antonio" to ULAN (500015183), VIAF (311436515 and 76586951) and Wikidata (Q8506). However, in Wikidata the URI is linked to the above VIAF record 76586951 and another VIAF record 125158790735238852393 (a duplicate). Also the VIAF record 311436515 is a duplicate, less complete and precise, but the Zeri photo archive did not discard any, eventually correct, match. This poses yet another problem in the selection of multiple candidate matches in an authority where we cannot, apparently, assume the uniqueness of records. In PHAROS we harmonise and store only the consistent links available.

The second case applies to photographers, who are first manually reconciled across institutions to co-design "umbrella terms" and share preferred forms of names. Secondly, entities flagged as people or ateliers are reconciled to Wikidata using the class as constraint, so as to eliminate false positive matches (e.g. a person reconciled to an organisation in Wikidata). Candidate matches are revised by humans. Umbrella terms and related entities are displayed in the interface as root names of hierarchical lists, used to filter artworks (Fig. 1).

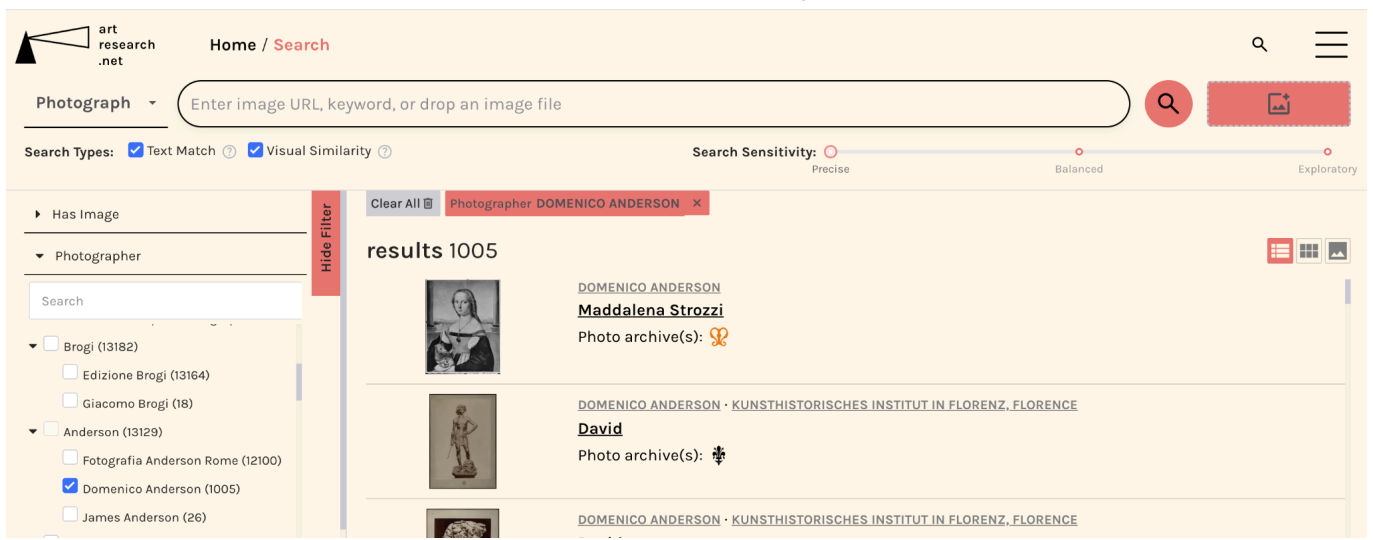

Figure 2. artreseach.net hierarchical facets (on the left) display umbrella terms grouping entities which could not find a one-to-one match due to their different ontological definition, despite being commonly known with a shared label.

Similar patterns exist across GLAM domains: RDA's "work group" mechanism, for instance, groups heterogeneous identities under a conceptual umbrella when authorship cannot be uniquely determined—mirroring our strategy for ambiguous photographers.

## 5.2 Structural and Semantic Ambiguity (Collections, Buildings, Places)

Similar strategies as those described above are used for reconciling keepers and places, which are automatically matched to Wikidata using string matching algorithms, and candidates are later reviewed by humans to check consistency in class declaration.

To address the ambiguity related to collections, PHAROS adopts collection-level description principles similar to those originating from (Heaney 2001). Our reconciliation process creates explicit nodes for collections, preserving institutional provenance while enabling alignment at different structural levels.

## 5.3 Granularity and Variant Traditions (Titles, Dates, Materials)

Differences in descriptive granularity—whether due to cataloguing policy, institutional tradition, or historical uncertainty—pose recurring reconciliation challenges. A photograph might be referenced with multiple titles (original, supplied, translated), a variable degree of description, or conflicting dates (for a positive: shot, printing, publication, reprint). Automatic reconciliation of such aspects across institutions or to third-party authorities is not desirable, therefore human curation is needed.

Rather than privileging a single "correct" value, PHAROS models all variants with explicit provenance, capturing their origin (institutional catalogue) and certainty type (e.g. traditional, attributed). Where heterogeneity obstructs navigation, we employ higher-level groupings. For instance, competing titles of relevant works are associated with a shared one, co-designed by institutions, when creating the link between two artwork records (further described below). Reviewers are asked to mark one of the existing titles as 'preferred' or, alternatively (if neither is convincing), to create a new one in English. Materials and techniques are manually reconciled to Getty AAT terms, so that experts are able to address the multi-layered semantics of those terms, and such taxonomical layers are later shown to users via hierarchical structures. Such an approach is similar to RDA's approach to work families and variant manifestations, which preserves detail while enabling users to browse coherent conceptual entities.

## 5.4 Organisational Governance and Workflow Distribution (artworks)

Reconciliation is not only a modelling challenge but a governance one. Institutions maintain their own cataloguing traditions, expectations of accuracy, and project priorities. Because no single authority file can be considered definitive, PHAROS adopts a distributed workflow: each archive validates candidate matches for its own records, and the combined taxonomy is then integrated into the shared knowledge graph.

One of the main goals is the record alignment across institutions to match equivalent artworks. PHAROS employs image-matching algorithms to propose cross-institutional reconciliations between artworks. Candidate matches are presented to experts through a user-friendly interface (Fig. 3), where they can be individually accepted or rejected. Unlike authority-based alignments—where candidates are drawn from a single external source—artworks can have multiple potential matches across different partner institutions, each requiring separate validation. The user interfaces developed for this purpose play a decisive role in how uncertainty is managed. By showing competing candidates, variant names, and provenance trails, the interface ensures transparency while still enabling curators to make informed decisions. These design choices reflect a broader reality recognised across Semantic Web technologies: under the Open World Assumption and the AAA principle ("Anyone can say Anything about Any thing"), the goal is not to impose absolute accuracy but to maintain a flexible, extensible, and provenance-aware representation. RDF's triple structure and the usage of named graphs supports this atomisation, allowing each institution's contribution to remain explicit and machine-readable.

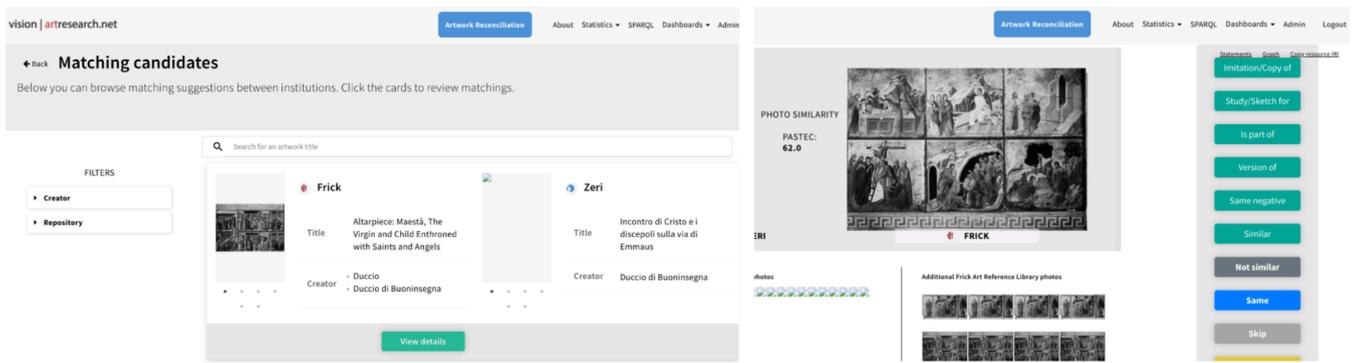

Figure 3. vision.artresearch.net is the artwork matching interface used by PHAROS partners. Reviewers can select pairs of artworks from different institutions and define the equivalence/associative relations among them.

Managing this reconciliation process raises two major socio-technical challenges: coordinating the division of work and handling the merging of records (together with all their inherited authority links).

Currently, seven institutions are piloting the artresearch.net infrastructure. Each is considered responsible for reconciling the entities referenced in its own records against third-party authorities. However, artwork-level matches often involve several institutions simultaneously, making it difficult to establish clear boundaries of responsibility. While an institution is naturally more inclined to review matches that include artworks from its own holdings, doing so also relieves other institutions from the shared workload of multi-institutional matches. To avoid inequities, the interface distributes an equal number of candidate matches to each institution, regardless of their origin. This groundwork sets the basis for future association members, who will build on these reconciliations by reviewing new matches involving their own artworks.

Once the cross-institutional reconciliation is in place for an artwork, the most prominent challenge derives from the actual merging of information and its presentation in a unified record view. All the pieces of information are associated with their provenance (in the interface summarised with the logo of the responsible institution), also those that are not affected by reconciliation in the first place. However, this operation requires further intervention by experts, who want to harmonise redundant text descriptions. For instance, a reconciled artwork inherits all the titles recorded by institutions, and a decision needs to be made in order to show final users only one title on which all institutions agree, instead of a list of (potentially long and very similar) titles. In PHAROS, reconciled artwork records offer both a main attributed title agreed between institutions and all the titles originally recorded, so as to preserve - although less prominently - institutional preferences and language variants. Other situations that deserve attention regard the hierarchical organisation of artefacts that are described separately but that all together form a coherent artwork (e.g. cycles of frescoes). In PHAROS, we preserve the records for each part of a composite work and we show the hierarchical relation in records. A similar solution is offered for the presentation of artwork subjects, which in PHAROS are reconciled to ICONCLASS. The taxonomic structure of ICONCLASS is preserved in records in order to show the different levels of granularity that institutions have adopted for the case at hand.

## 6. Discussion

Our case study shows that entity reconciliation in the cultural heritage domain—especially in art historical photo archives—cannot be treated as a purely technical task. It is a deeply interpretive and policy-driven process, shaped by the epistemic traditions and cataloguing practices of the contributing institutions. While algorithmic approaches, authority control, and Linked Open Data frameworks provide crucial infrastructure, they do not resolve the inherent ambiguities of cultural data. The PHAROS experience highlights that reconciliation requires embracing, rather than erasing, this ambiguity—while still producing usable, navigable knowledge graphs, which in turn requires the participation of domain experts in the decision-making process in several stages of the workflow.

A key insight is the tension between usability and philological correctness. Semantic Web technologies encourage creating very complex data, while, in practice, existing tooling is very restrictive and encourages flattening and normalisation of data so they can populate neat hierarchies, filters, and visualisations. Yet art historical data replete with vagueness, competing attributions, and evolving scholarly opinions. Strictly enforcing uniformity risks overwriting this complexity, while exposing it fully risks overwhelming users. PHAROS had to strike a compromise by creating umbrella terms, grouping uncertain or heterogeneous entities under shared labels, and using hierarchical user interface facets to conceal complexity by default but make it explorable on demand. These design choices reveal that reconciliation is as much about user experience as about data modelling.

The reconciliation challenges identified in this study are closely tied to the foundational assumptions of the Semantic Web. The Open World Assumption (OWA) ensures that absence of information cannot be interpreted as contradiction, making it possible to integrate partial, uncertain, or divergent statements without forcing premature closure. The AAA principle ("Anyone can say Anything about Any thing") helps explain why external authority datasets occasionally contain conflicting or incompatible assertions—an expected condition rather than an error. RDF's triple structure, designed for machine processing of atomic statements, enables fine-grained provenance capture but does not prescribe how statements should be presented to users. This creates a natural tension between modelling openness and interface-driven simplification, one that cultural heritage institutions must negotiate explicitly through reconciliation policies and interface design.

The project also demonstrates that fully automated reconciliation is not feasible given the data historical and institutional heterogeneity. Likewise, human review is not merely quality control but it is primarily epistemic work. Cultural heritage specialists do not accept probabilistic "best guesses" as sufficient, and expect human-informed, defensible assertions. This has implications for the role of AI and large language models (LLMs) in reconciliation. While LLMs and embeddings can help surface candidate matches or prune large sets of possibilities, they cannot replace expert judgment and should not be treated as sources of truth, especially in those cases where even humans would struggle to confirm identity (Doerr 2025). Their probabilistic nature clashes with the evidentiary expectations of humanists, and experiments showed that they can produce hallucinated links that undermine trust.

Another important point is that reconciliation is not a one-off task but an ongoing process. New data ingestions, evolving authority records, and shifting scholarly opinions continuously challenge existing alignments. This requires sustainable organisational strategies: iterative ingestion pipelines, versioning and provenance tracking of matches, and clear policies about when and how previously accepted links should be revised. These governance aspects are largely absent from the literature, yet they are crucial for long-term reliability and trust in aggregated cultural heritage platforms.

Ultimately, reconciliation must accept a degree of imperfection. Pursuing absolute precision is both unrealistic and counterproductive; what matters is to design processes that transparently manage uncertainty and make their compromises explicit. PHAROS illustrates that embracing many-to-many relationships, rather than forcing one-to-one or many-to-one matches, offers a viable path forward, provided that this complexity is mediated through thoughtful modelling and user interface design. This suggests that reconciliation should be framed less as a final goal of producing a single authoritative view, and more as an infrastructure to support the ongoing negotiation of meaning across heterogeneous cultural data.

# 7. Conclusion

This study has examined the reconciliation of heterogeneous data in art historical photo archives, focusing on the challenges and strategies developed within the PHAROS project. Our analysis has shown that reconciliation is not a purely technical task, but an interpretive and collaborative endeavour shaped by institutional legacies, evolving cataloguing practices, and disciplinary epistemologies. The absence of comprehensive authority lists, the presence of conflicting or incomplete metadata, and the intrinsic vagueness of art historical knowledge make full automation impossible and call for a careful balance between algorithmic support and expert-driven decision-making.

We have highlighted how strategies such as the creation of umbrella terms, the modelling of uncertainty, and the careful separation (or grouping) of entities allow us to preserve the historical specificity of cataloguing practices while still providing a coherent and usable interface to end users. This approach reframes reconciliation as an ongoing and revisable process rather than a one-time attempt to produce a definitive and authoritative dataset.

Rather than aspiring to full consistency across cultural heritage datasets—a goal associated with earlier interpretations of Universal Bibliographic Control and now widely recognised as neither realistic nor desirable—the emphasis today lies in achieving interoperable, provenance-aware representations that preserve plural voices. In this context, reconciliation is not a process of normalisation toward a single truth but a method for managing heterogeneity in ways that remain transparent and reusable. By adopting this perspective, PHAROS contributes to redefining reconciliation not as the eradication of heterogeneity, but as a framework to mediate and make sense of it, opening new possibilities for research on and through photographic archives. In future works, PHAROS will make available methods and data including reconciled entities openly via a bespoke OpenRefine service which will contribute to the broader art historical community with a service that inherits and serves the wealth of interpretive diversity recorded in art historical photo archives.

PHAROS is rooted in the domain of photographic archives. However, the reconciliation strategies and managerial challenges described here extend well beyond this context. Libraries face similar issues when aligning name authorities across LRM, RDA, and legacy cataloguing rules; archival institutions encounter comparable ambiguity in EAC-CPF records and biographical entities; museums must reconcile heterogeneous object descriptions, attributions, and collection histories across systems such as ULAN, AAT, and local thesauri. The framework proposed here—integrating identification, modelling, workflow, and interface strategies—therefore offers a reusable structure for any GLAM organisation seeking to consolidate heterogeneous datasets under a shared, semantically interoperable environment.

# Acknowledgments

The authors would like to thank all colleagues who contributed to the manual reconciliation work and to the discussions that informed the decision-making process. In particular, we acknowledge the careful reviews and constructive input provided by Ute Dercks, Dagmar Keultjes, Kurt Scharenberg (KHI); Pietro Liuzzo (Hertziana); John McQuaid, Kerri A. Pfister (Frick); Spyros Koulouris (I Tatti); Christoph Glorious (Marburg) and Louisa Wood Ruby, whose expertise and attention to detail were essential in group discussions and in resolving discrepancies and refining final decisions.